\documentclass[9pt,twocolumn,twoside]{osajnl_mod}
\usepackage{mathrsfs}

\journal{josab_vanilla} 

\setboolean{shortarticle}{false} 

\title{Photorefractive second-harmonic detection for ultrasound-modulated
optical tomography}

\author[1]{Steven J. Byrnes}

\affil[1]{The Charles Stark Draper Laboratory, Inc., 555 Technology Sq., Cambridge, Massachusetts, USA}

\affil[*]{Corresponding author: sbyrnes@draper.com}

\ociscodes{(170.1065) Acousto-optics; (070.1060) Acousto-optical signal processing; (190.5330) Photorefractive optics.}

\begin{abstract}
Ultrasound-modulated optical tomography enables sharp 3D optical imaging
of tissues and other turbid media, but the light modulation signals
are hard to sensitively measure. A common solution, involving photorefractive
crystals, enables the measurement of a relatively slow and low-spatial-resolution
signal tracking the envelope of the ultrasound wave. We reexamine
the photorefractive detection principle both intuitively and quantitatively,
and from this analysis we predict that the photodetector should additionally
see a fast response at twice the ultrasound frequency, and correspondingly
high spatial frequency. The fast and slow response usually have similar
amplitudes and reveal complementary information, thus allowing ultrasound-modulated
optical tomography to create dramatically sharper tomographic images
under the same measurement conditions, integration time, and experimental
complexity. 
\end{abstract}

\setboolean{displaycopyright}{false}

\begin{document}

\maketitle

\section{Introduction}\label{sec:Introduction}

Ultrasound-modulated optical tomography \cite{gunther2017reviewof},
also called acousto-optic tomography or acousto-optic imaging, is
an emerging medical imaging technology promising a uniquely appealing
combination of features, including high spatial resolution ($<1\text{mm}{}^{3}$),
deep penetration ($\gg1\text{cm}$), sensitivity to visible texture
(light scattering and refractive index contrast, unlike photoacoustic
tomography which is limited to light absorption contrast), excellent
safety profile (non-invasive, no contrast agents, no ionizing radiation),
and a path to low-cost, even portable, equipment.

In ultrasound-modulated optical tomography, one sends laser light
(typically near-infrared) and ultrasound simultaneously into a tissue.
The two waves interact inside the tissue (Sec.~\ref{subsec:IndexVsDisplacement}
below), with the effect that the light exiting the tissue is amplitude-modulated
and phase-modulated at the ultrasound frequency. The amount of light
modulation is the signal to be measured in this technique. This modulation
signal is especially useful because it inherits the spatial resolution
of ultrasound\textemdash which for example can be focused to small
points deep inside the tissue and scanned around\textemdash while
still maintaining the image contrast of light, including distinguishing
subtle properties of soft tissues without contrast agents, and measuring
blood volume, flow, and oxygenation. Light by itself (diffuse optical
tomography) has comparatively poor spatial resolution because of scattering
by the tissue\textemdash to get appreciable depth, it is often necessary
to use photons that have been scattered hundreds of times inside the
tissue.

One of the principal challenges in implementing ultrasound-modulated
optical tomography is detecting the light modulation signal. Unfortunately,
light exits the tissue in a diffuse, high-\'{e}tendue, laser speckle pattern.
Thus the most naive approach\textemdash collecting lots of light into
a large-area fast detector, and watching its output for modulation at
the ultrasound frequency\textemdash works quite poorly. There are
many speckles entering the detector, and they are modulated with different
phases, with some getting brighter at the same time as others are
getting dimmer, and thus the speckles largely cancel each other out.

To work around this issue, a wide variety of alternative approaches
have been explored, including small detectors \cite{wang1995continuouswave},
CCD or CMOS imaging detectors \cite{lev^eque1999ultrasonic}, narrowband
filters \cite{sakadvzic2004highresolution,li2008detection}, and many
others \cite{gunther2017reviewof}. However, challenges remain in
combining very high detector \'{e}tendue (i.e. collecting many photons), reasonable
cost and convenience, high SNR, and compatibility with fast, high-bandwidth
measurements, such as in Ref.~\cite{laudereau2016ultrafast}.

\begin{figure}
\begin{centering}
\includegraphics[width=0.4\textwidth]{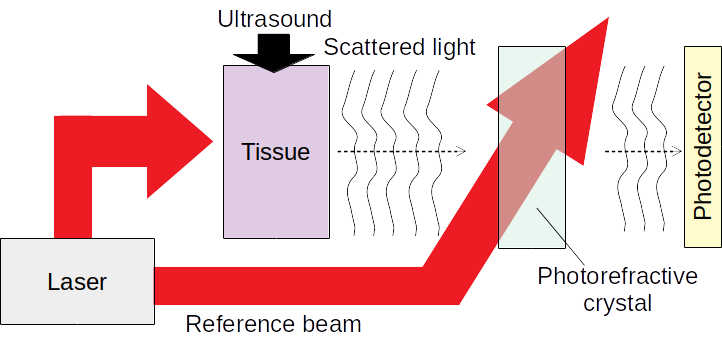}
\par\end{centering}
\caption{(a) Schematic block diagram of the photorefractive detection system.
A CW laser illuminates a tissue, creating a speckle pattern of scattered
light. This is overlapped with a reference beam in a photorefractive
crystal, and then a photodetector integrates the total light power
received.\label{fig:BlockDiagram}}
\end{figure}

One of the most promising and widely-used detection methods is photorefractive
detection \cite{murray2004detection,gross2005theoretical,gross2009detection,lai2012ultrasoundmodulated},
as shown in Fig.~\ref{fig:BlockDiagram}. In this technique, the
outgoing speckle pattern passes through a photorefractive crystal.
Meanwhile a ``reference beam'' light source passes through the same
photorefractive crystal. In the configuration we discuss in this paper,
the reference beam is at the same frequency as the laser illuminating
the tissue. (This is not the only possible photorefractive detection
configuration \cite{gross2005theoretical,lesaffre2009acoustooptical},
but it is most appropriate for fast, high-bandwidth measurements.)
As discussed below, the reference beam and the speckle pattern undergo
two-wave mixing, and this leads to a signal indicating the amount
of ultrasound modulation. In photorefractive detection, unlike the
naive approach mentioned above, the different speckles do \emph{not}
cancel each other out, and therefore a large-area detector can be
used without reducing the relative strength of the modulation signal.

\begin{figure}
\centering{}\includegraphics[width=0.3\textheight]{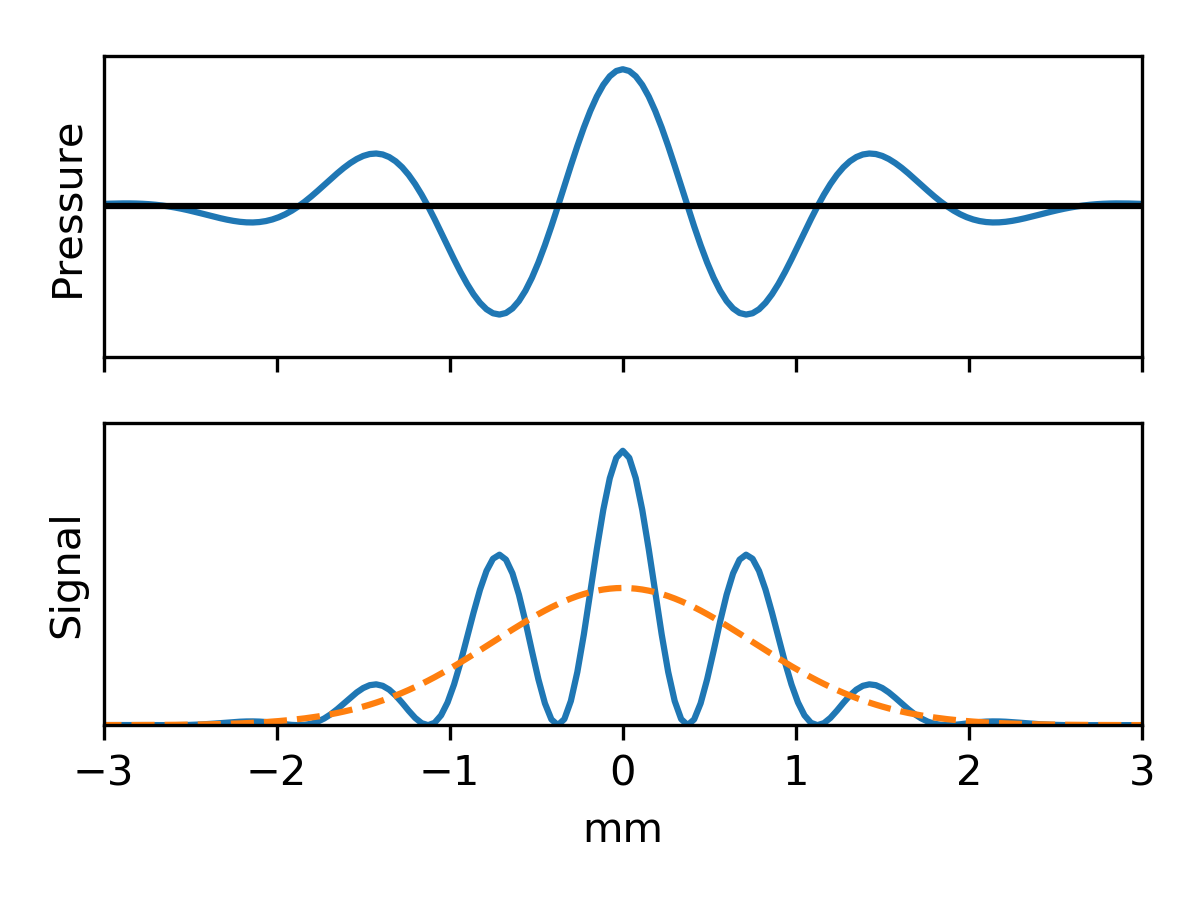}\caption{Top: An example ultrasound waveform, showing a snapshot of a short pulse at 1MHz
with $v_{\text{sound}}=1.5\text{km}/\text{s}$, as it travels through the medium. Bottom: If this wave
is used for ultrasound-modulated optical tomography with photorefractive
detection, under typical operating conditions (Sec.~\ref{subsec:IndexVsDisplacement}),
we will argue that the spatial sensitivity at each moment is approximately
as shown by the solid blue line. If the detector is too slow to measure
the fast oscillations, the spatial sensitivity would instead be the
dashed line, as in previous work.\label{fig:Pulses}}
\end{figure}

In previous work \cite{sui2005imaging,gross2005theoretical,laudereau2016ultrafast},
photorefractive detection has been treated as having a spatial resolution
equal to the envelope of the ultrasound wave. We will argue that if
a faster optical detector is used (much faster than the ultrasound frequency), then the spatial resolution has much finer structure,
as in Fig.~\ref{fig:Pulses}, associated with oscillations at the
second harmonic of the ultrasound frequency. In many situations, these signals are beneficial to measure: they
can reveal very small features and sharp edges (see Fig.~\ref{fig:Pulses}); higher-frequency signals
are often easier to detect over background; higher-frequency signals
offer more data channels per unit time; the fast and slow signals
can be measured simultaneously with little extra effort, enabling
measurement of both large-scale and small-scale structure; and for
technical reasons discussed below, the high-frequency signal can sometimes
have a somewhat different image contrast than the low-frequncy signal,
such that we can  make certain features stand out (Sec.~\ref{sec:Conclusion}
below).

\subsection{Nonlinear modulation vs. nonlinear detection\label{subsec:Nonlinear-modulation-vs.}}

Photorefractive detection, as we describe it here, involves the \emph{nonlinear}
detection of a \emph{linear} acousto-optic modulation signal, and
we will find that the second-harmonic signal discussed in this paper
appears in the lowest order of perturbation theory. This topic should
thus be distinguished from \emph{nonlinear modulation} of light in
the tissue. Examples of the latter include purely acoustic second-harmonic
generation\textemdash when the acoustic amplitude is sufficiently
strong that the tissue's mechanical response is anharmonic\textemdash or
acousto-optic second-harmonic generation\textemdash when the acousto-optic
interaction is sufficiently strong that the light modulation is anharmonic.
Example of nonlinear modulation include Refs.~\cite{selb2002nonlinear,ruan_pulse_2012},
in which second-harmonic signals in ultrasound-modulated optical
tomography were seen in CCD-based measurements.

Nonlinear modulation offers intriguing potential applications, particularly
with focused ultrasound, but is weaker than the main signal at the
ultrasound frequency. Photorefractive detection offers an exciting
avenue because, regardless of ultrasound intensity, we predict a second-harmonic
signal with similar amplitude to the slow envelope signal\textemdash and
experience shows that the latter can be measured even at high speeds
with unfocused beams \cite{laudereau2016ultrafast}.

To the extent that nonlinear modulation occurs, a photorefractive
detector would see it as a \emph{fourth} harmonic signal.

\section{Qualitative explanation}

\subsection{Index effect vs. displacement effect\label{subsec:IndexVsDisplacement}}

We predict that the strength and presence of the second-harmonic signal
depends critically on the phase relation between an ultrasound wave
and the optical modulation it induces. Having a consistent, predictable
phase relation is critical to ensuring that the signal adds coherently,
rather than randomly, over all the different light paths meandering
through the highly-scattering tissue.

Previous work \cite{wang2001mechanisms,wang2001mechanisms2} has identified
two mechanisms by which ultrasound modulates light, each with a different
phase relation. The first mechanism is the \emph{index effect}, wherein
the pressure of the sound wave changes the refractive index of the
water in the tissue (piezo-optic effect), which in turn changes the
optical phase of light passing through that region of tissue. The
second is the \emph{displacement effect}, where the light-scattering
structures in the tissue move back and forth at the ultrasound frequency.

\begin{figure*}
\centering{}\includegraphics[width=0.65\textwidth]{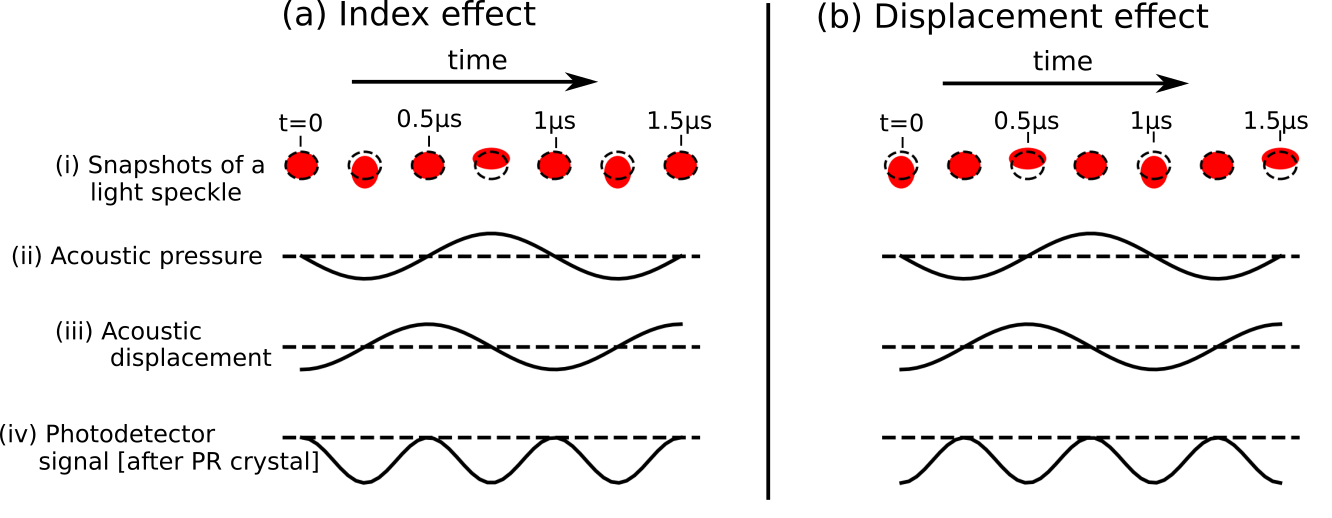}\caption{Schematic depiction of speckle modulation during 1.5 ultrasound cycles,
with ultrasound frequency 1MHz, showing (i) a series of snapshots of a representative laser
speckle exiting the tissue at different times, (ii) the acoustic pressure and (iii) displacement
at the part of the tissue causing this modulation, and (iv) the photodetector
signal after the photorefractive crystal. In all cases, dashed lines
indicate the equilibrium with no ultrasound. (a) If the light modulation
is caused by the index effect (Sec.~\ref{subsec:IndexVsDisplacement}),
the light speckle changes in phase with the pressure\textemdash i.e.
when pressure is at its equilibrium value, the speckle is in its equilibrium
position; (b) If the light modulation is instead caused by the displacement
effect, the light speckles change in phase with the displacement.
Either way, as explained in Sec.~\ref{subsec:CS}, the signal (iv)
shrinks when the speckles (i) are displaced from their time-averaged
configuration.\label{fig:SpeckleDance}}
\end{figure*}

As shown in Fig.~\ref{fig:SpeckleDance}, the two mechanisms lead
to different optical modulation phases, and indeed, if both are present
equally, they tend to create equal and opposite second-harmonic signals
that cancel out in a photorefractive detection system.

However, for typical tissue parameters and ultrasound frequencies,
the two mechanisms are \emph{not }present equally; instead, the index
effect predominates over the displacement effect \cite{wang2001mechanisms,wang2001mechanisms2}.
For example, for tissue with light scattering coefficient $1\text{mm}^{-1}$,
the index effect is predicted to account for $\frac{2}{3}$ of the
total light modulation with 500kHz ultrasound, $\frac{3}{4}$ with
1MHz ultrasound, and increasing towards 100\% at higher frequency.
(More generally, for water-based media, the index effect predominates
by at least a 2:1 ratio on condition that the ultrasound wavelength
$\lambda_{\text{US}}<2l$, where $l$ is the mean free path for photon scattering~\cite{wang2001mechanisms,wang2001mechanisms2}.)

\subsection{Two-wave mixing\label{subsec:Two-wave-mixing}}

A photorefractive crystal has the defining property that the crystal's
refractive index changes in response to how much light intensity is
at any given point. This gives rise to ``two-wave mixing'' \cite{delaye1995transmission,chi2009ageneral,gross2005theoretical,yeh_introduction_1993}.
In this phenomenon, two light beams of the same wavelength shine into
the crystal. The interference pattern between these two beams produces
peaks and nulls of intensity within the crystal, which in turn creates
a volumetric diffraction grating in the crystal. This grating diffracts
each beam into the spatial mode of the other beam. In this case, one
of the ``beams'' is a complicated laser speckle pattern, but the
principle still works: the volume grating diffracts light from the
reference beam into the complicated laser speckle spatial profile,
and vice-versa.

\begin{figure}
\begin{centering}
\includegraphics[width=0.4\textwidth]{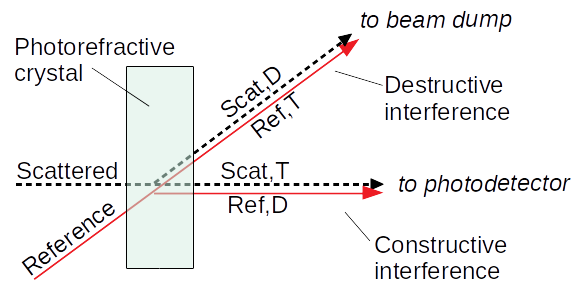}\caption{Zoomed-in diagram of the photorefractive crystal of Fig.~\ref{fig:BlockDiagram}.
The four beams exiting the crystal on the right are, from top to bottom,
the diffracted scattered light, the transmitted reference beam, the
transmitted scattered light, and the diffracted reference beam. The
labels ``destructive interference'' and ``constructive interference''
are as discussed in Sec.~\ref{subsec:Two-wave-mixing}. \label{fig:BlockDiagramZoom}}
\par\end{centering}
\end{figure}

Referring to Fig.~\ref{fig:BlockDiagramZoom}, there is interference
between the diffracted reference beam and the transmitted scattered
light traveling to the photodetector, and there is also interference
between the transmitted reference beam and diffracted scattered light
traveling to the beam dump. For definiteness, we will assume that
the former interference is constructive, and hence (by conservation
of energy) that the latter interference is destructive. This case
corresponds to positive gain for the scattered light, i.e. transfer
of power from the reference to the scattered light paths. In general,
the sign of gain, and correspondingly the choice of which of the beams
is amplified and which is depleted, depends on the detailed mechanism
of photorefraction, including the crystallographic axis, beam propagation
directions, and direction of the electric field applied across the crystal (if any) \cite{garrett1992highbeamcoupling}.
If we instead made the opposite assumption, i.e. transfer of power
\emph{into} the reference beam, the system would still work, but some
signs would be flipped in the discussion below, e.g. the presence
of ultrasound would tend to increase, rather than decrease, the photodetector
signal.

What makes two-wave mixing particularly useful for our purposes is
that the crystal's photorefractive response time, $\tau_{\text{PR}}$,
is much slower than the ultrasound frequency. Therefore the diffracted
reference beam is not quite a copy of the speckle pattern exiting
the tissue, but rather a \emph{time-averaged copy} of that speckle
pattern. This is shown schematically in Fig.~\ref{fig:SpeckleDance}(i),
where a representative speckle exiting the tissue (red dot) is modulated,
while the corresponding diffracted speckle from the reference beam
(dashed line) stays stationary. The time-averaged copy enables ``change
detection'' for the light: We are essentially comparing the speckle
pattern to its time-averaged copy, and if they are very different,
we infer that the light is being strongly modulated by the ultrasound.

\subsection{Mode overlap determines output signal\label{subsec:CS}}

A key aspect of the photorefractive detection method is the interference
between the transmitted scattered light and the diffracted reference
beam. Each beam by itself has an approximately fixed total power,
but to the extent that the two beams interfere, it sends more power
to the photodetector and less power to the beam dump (Fig.~\ref{fig:BlockDiagramZoom}).

If both beams are static for a long time compared to $\tau_{\text{PR}}$,
then a properly-configured photorefractive crystal will naturally
develop an index profile that maximizes interference, such that the
beams are maximally in phase and overlapped. This maximizes the power
to the photodetector, and minimizes the signal to the beam dump of
Fig.~\ref{fig:BlockDiagramZoom}. However, if the scattered light
is modulated by the ultrasound, its phase and amplitude profile will
not perfectly match the diffracted reference beam, so the interference
will be less effective, and the photodetector will see less light.
This phenomenon is part of Fig.~\ref{fig:SpeckleDance}: the more
the speckles are altered from their time-averaged pattern (Fig.~\ref{fig:SpeckleDance}(i)),
the less power is detected (Fig.~\ref{fig:SpeckleDance}(iv)).

As stated above, two interfering waves convey maximum power when they
are in phase and have matching spatial profiles. This familiar fact
from wave mechanics can be mathematically proven by writing \cite{gross2005theoretical}:
\begin{align}
\text{Photodetector signal} &=\int d^{2}\mathbf{r}\left|\mathscr{E}_{S}(\mathbf{r})+\mathscr{E}_{D}(\mathbf{r})\right|^{2} \nonumber \\
&=C+2\int d^{2}\mathbf{r}\text{Re}\left[\mathscr{E}_{S}^{*}(\mathbf{r})\mathscr{E}_{D}(\mathbf{r})\right]\label{eq:ES_plus_ED}
\end{align}
where $\text{Re}\left[\mathscr{E}_{S}(\mathbf{r})e^{-i\omega_{0}t}\right]$
is the electric field of the transmitted scattered light in Fig.~\ref{fig:BlockDiagramZoom};
$\text{Re}\left[\mathscr{E}_{D}(\mathbf{r})e^{-i\omega_{0}t}\right]$
is the electric field of the diffracted reference beam in Fig.~\ref{fig:BlockDiagramZoom};
and $C\equiv\int d^{2}\mathbf{r}\left|\mathscr{E}_{S}(\mathbf{r})\right|^{2}+\int d^{2}\mathbf{r}\left|\mathscr{E}_{D}(\mathbf{r})\right|^{2}$
is a constant approximately independent of the presence or absence
of ultrasound. Now we invoke the Cauchy\textendash Schwarz inequality:
\begin{subequations}
\begin{multline}
\int d^{2}\mathbf{r}\text{Re}\left[\mathscr{E}_{S}^{*}(\mathbf{r})\mathscr{E}_{D}(\mathbf{r})\right]=\sqrt{\int d^{2}\mathbf{r}\left|\mathscr{E}_{S}^{2}(\mathbf{r})\right|\cdot \int d^{2}\mathbf{r}\left|\mathscr{E}_{D}^{2}(\mathbf{r})\right|} \\
\text{if }\mathscr{E}_{S}\text{ \& }\mathscr{E}_{D}\text{ have the same phase \& spatial mode}
\end{multline}
\begin{multline}
\int d^{2}\mathbf{r}\text{Re}\left[\mathscr{E}_{S}^{*}(\mathbf{r})\mathscr{E}_{D}(\mathbf{r})\right]<\sqrt{\int d^{2}\mathbf{r}\left|\mathscr{E}_{S}^{2}(\mathbf{r})\right|\cdot \int d^{2}\mathbf{r}\left|\mathscr{E}_{D}^{2}(\mathbf{r})\right|} \\ \text{otherwise}
\end{multline}\label{eq:C-S}
\end{subequations}
If $\mathscr{E}_{S}$ and $\mathscr{E}_{D}$ have the same phase profile
and spatial mode, i.e. if $\frac{\mathscr{E}_{S}(\mathbf{r})}{\mathscr{E}_{D}(\mathbf{r})}$
is the same positive constant everywhere, then the photodetector signal
is maximized. If $\mathscr{E}_{S}$ and $\mathscr{E}_{D}$ are mis-matched
at all, the photodetector signal goes down.

As shown in Fig.~\ref{fig:SpeckleDance}(iv), the presence of ultrasound
both decreases the time-averaged photodetector signal, and causes
a fast modulation at twice the ultrasound frequency. The former is
the familiar slow signal whose spatial sensitivity tracks the envelope
of the ultrasound wave; the latter is the subject of this article.

\section{Quantitative derivation}

As above, we say $\omega_{0}$ is the laser frequency and $\omega_{\text{US}}$
is the ultrasound frequency (assumed monochromatic for simplicity).
We write the scattered field (i.e. the speckle pattern coming out
of the tissue) as $\text{Re}\left(\mathscr{E}_{S}(\mathbf{r}_{e},t)e^{-i\omega_{0}t}\right)$,
where $\mathscr{E}_{S}$ is a complex amplitude and $\mathbf{r}_{e}$
is any point in the exterior of the tissue where the light exits as
a speckle pattern. It will be most convenient to set $\mathbf{r}_{e}$
to points on the photodetector surface, in which case $\mathscr{E}_{S}$
incorporates the effects of passing the photorefractive crystal, i.e.
some absorptive and diffractive loss, both usually small in practice
\cite{gross2005theoretical}. To simplify the descriptions, we neglect
speckle decorrelation, i.e. we assume that $\mathscr{E}_{S}(\mathbf{r}_{e},t)$
depends on time \emph{only} because of the presence of ultrasound,
and not because of blood flow or other motion. (Photorefractive detection
continues to work in the presence of speckle decorrelation, but the
signal strength goes down when the speckle decorrelation time is shorter
than the photorefractive response time $\tau_{\text{PR}}$. In practice, sub-millisecond
values of $\tau_{\text{PR}}$ are known to be feasible \cite{lesaffre2006submillisecond},
and this is sufficiently low for most \emph{in vivo} applications.)

The ultrasound wave inside the tissue creates local pressure change
$P(\mathbf{r}_{i},t)$ and local displacement $D(\mathbf{r}_{i},t)$,
where $t$ is time and $\mathbf{r}_{i}$ is a point inside the tissue.
For example, a plane wave would have $P(\mathbf{r}_{i},t)=P_{0}\cos\left(\mathbf{k}\cdot\mathbf{r}_{i}-\omega_{\text{US}}t\right)$,
$D(\mathbf{r}_{i},t)=\frac{P_{0}}{Z_{\text{US}}\omega_{\text{US}}}\sin\left(\mathbf{k}\cdot\mathbf{r}_{i}-\omega_{\text{US}}t\right)$,
where $Z_{\text{US}}$ is the specific acoustic impedance (note that
the displacement is $90^{\circ}$ out of phase with the pressure change).
Then the effect of the ultrasound can be written using Green's functions
as:

\begin{align}
\mathscr{E}_{S}(\mathbf{r}_{e},t)&=\mathscr{E}_{S,0}(\mathbf{r}_{e},t) \nonumber \\
&\; +\, \int d^{3}\mathbf{r}_{i}G_{n}(\mathbf{r}_{i},\mathbf{r}_{e})P(\mathbf{r}_{i},t)+\int d^{3}\mathbf{r}_{i}G_{d}(\mathbf{r}_{i},\mathbf{r}_{e})D(\mathbf{r}_{i},t)\label{eq:FirstGreen}
\end{align}
where $\mathscr{E}_{S,0}$ is the field in the absence of ultrasound;
$G_{n}$ is the index-effect Green's function, defined such that a
unit change of pressure at the point $\mathbf{r}_{i}$ in the tissue
causes an optical field amplitude change of $G_{n}(\mathbf{r}_{i},\mathbf{r}_{e})$
in the speckle pattern, at the point $\mathbf{r}_{e}$ external to
the tissue; and $G_{d}$ is the displacement-effect Green's function
defined analogously. Eq.~(\ref{eq:FirstGreen}) is a first-order
approximation, i.e. assuming that the ultrasound only modestly changes
the light flow patterns. This is usually reasonable, and see Sec.~\ref{subsec:Nonlinear-modulation-vs.}
for further discussion. As mentioned above, we are ignoring blood
flow and other motion, so $G_{n}$ and $G_{d}$ do not depend on time.

If we did \emph{not} have a photorefractive crystal and reference
beam, but instead just sent the scattered light into a single-pixel
large-area photodetector, it would register an intensity fluctuation
of:
\[
\begin{aligned}\Delta I(t) & \equiv\int d^{2}\mathbf{r}_{e}\left(\left|\mathscr{E}_{S}(\mathbf{r}_{e},t)\right|^{2}-\left|\mathscr{E}_{S,0}(\mathbf{r}_{e})\right|^{2}\right)\\
 & =\int d^{2}\mathbf{r}_{e}\left(2\text{Re}\left[\mathscr{E}_{S,0}^{*}(\mathbf{r}_{e})\int d^{3}\mathbf{r}_{i}\left(G_{n}(\mathbf{r}_{i},\mathbf{r}_{e})P(\mathbf{r}_{i},t)\right)\right]\right)\\
 & \quad+\int d^{2}\mathbf{r}_{e}\left(2\text{Re}\left[\mathscr{E}_{S,0}^{*}(\mathbf{r}_{e})\int d^{3}\mathbf{r}_{i}\left(G_{d}(\mathbf{r}_{i},\mathbf{r}_{e})D(\mathbf{r}_{i},t)\right)\right]\right)\\
 & \quad+\int d^{2}\mathbf{r}_{e}\left|\int d^{3}\mathbf{r}_{i}\left(G_{n}(\mathbf{r}_{i},\mathbf{r}_{e})P(\mathbf{r}_{i},t)+G_{d}(\mathbf{r}_{i},\mathbf{r}_{e})D(\mathbf{r}_{i},t)\right)\right|^{2}
\end{aligned}
\]
As discussed in Sec.~\ref{sec:Introduction}, this fluctuation $\Delta I$
is quite small: The ultrasound slightly changes photon phases and
paths, but does not systematically change the total flux of photons
exiting the tissue into the large detector area. (Indeed, if this
modulation were not so small, there would be no need for the photorefractive
crystal!)

As described in \cite{gross2005theoretical} (see also Eq.~(\ref{eq:ES_plus_ED})),
under typical experimental conditions, the photodetector measures
a quantity proportional to:
\begin{equation}
\text{Signal}(t)=\int d^{2}\mathbf{r}_{e}\left(2\text{Re}\left[\mathscr{E}_{S,0}^{*}(\mathbf{r}_{e})\mathscr{E}_{S}(\mathbf{r}_{e},t)\right]\right)\label{eq:SignalStart}
\end{equation}
where $\mathscr{E}_{S,0}$ is proportional to the diffracted reference
beam (Fig.~\ref{fig:BlockDiagramZoom}), which imitates a time-averaged
version of the scattered light speckle pattern $\mathscr{E}_{S}$
(Sec.~\ref{subsec:Two-wave-mixing}).

In Eq.~(\ref{eq:SignalStart}), there is a time-independent and ultrasound-independent
offset of $2\int\left|\mathscr{E}_{S,0}\right|^{2}$, which can be
experimentally subtracted off by, for example, comparing the signal
with and without ultrasound \cite{gross2005theoretical}. Discarding
that, we are left with:
\begin{multline*}
\text{Signal}(t) = \Delta I(t)-\int d^{2}\mathbf{r}_{e}\left|\int d^{3}\mathbf{r}_{i}\left(G_{n}(\mathbf{r}_{i},\mathbf{r}_{e})P(\mathbf{r}_{i},t) \right. \right. \\ \left.  + G_{d}(\mathbf{r}_{i},\mathbf{r}_{e})D(\mathbf{r}_{i},t)\right)\Big|^{2}
\end{multline*}
To simplify this, first we ignore the small quantity $\Delta I(t)$,
then we note that the quantities $G_{n}(\mathbf{r}_{i},\mathbf{r}_{e})$
and $G_{d}(\mathbf{r}_{i}',\mathbf{r}_{e}')$ are statistically uncorrelated
(have a random phase relation). This leads to:
\begin{multline*}
\left\langle \text{Signal}(t)\right\rangle =\left\langle -\int d^{2}\mathbf{r}_{e}\left|\int d^{3}\mathbf{r}_{i}G_{n}(\mathbf{r}_{i},\mathbf{r}_{e})P(\mathbf{r}_{i},t)\right|^{2} \right. \\ \left. - \int d^{2}\mathbf{r}_{e}\left|\int d^{3}\mathbf{r}_{i}G_{d}(\mathbf{r}_{i},\mathbf{r}_{e})D(\mathbf{r}_{i},t)\right|^{2}\right\rangle 
\end{multline*}
Next we try to simplify: 
\begin{multline*}
\left|\int d^{3}\mathbf{r}_{i}G_{n}(\mathbf{r}_{i},\mathbf{r}_{e})P(\mathbf{r}_{i},t)\right|^{2} \\
=\int d^{3}\mathbf{r}_{i}\int d^{3}\mathbf{r}_{i}'G_{n}(\mathbf{r}_{i},\mathbf{r}_{e})P(\mathbf{r}_{i},t)G_{n}^{*}(\mathbf{r}_{i}',\mathbf{r}_{e})P(\mathbf{r}_{i}',t)
\end{multline*}
Under typical circumstances, the statistical correlation between $G_{n}(\mathbf{r}_{i},\mathbf{r}_{e})$
and $G_{n}(\mathbf{r}_{i}',\mathbf{r}_{e})$ smoothly decays to zero
with increasing $\left|\mathbf{r}_{i}-\mathbf{r}_{i}'\right|$, which
leads to:
\[
\left\langle \left|\int d^{3}\mathbf{r}_{i}G_{n}(\mathbf{r}_{i},\mathbf{r}_{e})P(\mathbf{r}_{i},t)\right|^{2}\right\rangle =B_{n}\int d^{3}\mathbf{r}_{i}\left\langle \left|G_{n}(\mathbf{r}_{i},\mathbf{r}_{e})\right|^{2}\right\rangle P^{2}(\mathbf{r}_{i},t)
\]
where $B_{n}=\frac{\int d^{3}\mathbf{r}_{i}'P(\mathbf{r}_{i}')\left\langle G_{n}(\mathbf{r}_{i},\mathbf{r}_{e})G_{n}^{*}(\mathbf{r}_{i}',\mathbf{r}_{e})\right\rangle }{P(\mathbf{r}_{i})\left\langle \left|G_{n}(\mathbf{r}_{i},\mathbf{r}_{e})\right|^{2}\right\rangle }$
is roughly the ratio of ultrasound pressure averaged over a sphere
centered at a point to pressure at the center of that sphere, where
the sphere size corresponds to the autocorrelation length of $G_{n}$
mentioned above. If the ultrasound waveform is approximately sinusoidal,
$B_{n}$ will be approximately the same real constant at each point
$\mathbf{r}_{i}$, although the constant will vary with ultrasound
frequency, scattering coefficient, and other parameters. The $G_{d}$
case is similar, so the end result is:
\begin{multline*}
\left\langle \text{Signal}(t)\right\rangle \propto-\int d^{2}\mathbf{r}_{e}\int d^{3}\mathbf{r}_{i}\left(B_{n}\left\langle \left|G_{n}(\mathbf{r}_{i},\mathbf{r}_{e})\right|^{2}\right\rangle P^{2}(\mathbf{r}_{i},t)\right. \\ \left. + \, B_{d}\left\langle \left|G_{d}(\mathbf{r}_{i},\mathbf{r}_{e})\right|^{2}\right\rangle D^{2}(\mathbf{r}_{i},t)\right)
\end{multline*}
Define the \emph{total modulation} $M$ ($M_{n}$ for index effect
and $M_{d}$ for displacement effect) by 
\begin{gather*}
M_{n}(\mathbf{r}_{i})\equiv B_{n}\int d^{2}\mathbf{r}_{e}\left\langle \left|G_{n}(\mathbf{r}_{i},\mathbf{r}_{e})\right|^{2}\right\rangle \\
M_{d}(\mathbf{r}_{i})\equiv B_{d}\int d^{2}\mathbf{r}_{e}\left\langle \left|G_{d}(\mathbf{r}_{i},\mathbf{r}_{e})\right|^{2}\right\rangle 
\end{gather*}
Mapping these quantities is an end-goal of ultrasound-modulated optical
tomography, as they reveal information about how much light passes
through each point $\mathbf{r}_{i}$, how much light scattering is
happening at that point, how much water content is at that point,
and so on. We have:
\[
\left\langle \text{Signal}(t)\right\rangle \propto-\int d^{3}\mathbf{r}_{i}\left(M_{n}(\mathbf{r}_{i})P^{2}(\mathbf{r}_{i},t)+M_{d}(\mathbf{r}_{i})D^{2}(\mathbf{r}_{i},t)\right)
\]
Finally, we take the simple example of a monochromatic ultrasound
plane wave in a uniform infinite medium: $P(\mathbf{r}_{i},t)=P_{0}\cos\left(\mathbf{k}\cdot\mathbf{r}_{i}-\omega_{\text{US}}t\right)$
and $D(\mathbf{r}_{i},t)=\frac{P_{0}}{Z_{\text{US}}\omega_{\text{US}}}\sin\left(\mathbf{k}\cdot\mathbf{r}_{i}-\omega_{\text{US}}t\right)$.
We find that our signal is related to the Fourier coefficients of
$M$:
\begin{align}
\frac{\left\langle \text{Signal}(t)\right\rangle }{P_{0}^{2}} &\propto-\left(\tilde{M}_{n}(0)+\frac{1}{Z_{\text{US}}^{2}\omega_{\text{US}}^{2}}\tilde{M}_{d}(0)\right)\nonumber\\
&\; -\, \text{Re}\left[\left(\tilde{M}_{n}(2\mathbf{k})-\frac{1}{Z_{\text{US}}^{2}\omega_{\text{US}}^{2}}\tilde{M}_{d}(2\mathbf{k})\right)e^{-2i\omega_{\text{US}}t}\right]\label{eq:final}
\end{align}
The first term on the right side is more generally the difference-frequency
term, which tracks the envelope of the ultrasound wave and has been
frequently measured in the literature. The second term is the sum-frequency
term, which measures higher spatial frequency, and appears at higher temporal
frequency (up to twice the highest ultrasound frequency). In the usual circumstance that $M_{n}\gg \frac{1}{Z_{\text{US}}^{2}\omega_{\text{US}}^{2}}M_{d}$ (see Sec.~\ref{subsec:IndexVsDisplacement}), the fast term and slow term have inherently equal amplitudes, though the actual amplitude of each will depend on the amplitude of the corresponding Fourier component of the image. For example, in a tissue with no fine features or sharp edges (where ``fine'' and ``sharp'' are compared to the ultrasound wavelength), the fast signal would be much weaker than the slow signal; conversely, a tissue with large-scale uniformity but small-scale random texture would generally show a stronger fast signal than slow signal. On the other hand, when $M_{n}\approx\frac{1}{Z_{\text{US}}^{2}\omega_{\text{US}}^{2}}M_{d}$, as when the ultrasound wavelength is much larger than the optical scattering mean free path (see Sec.~\ref{subsec:IndexVsDisplacement}), then the fast signal is expected to be much weaker than the slow signal under most circumstances.

\section{Conclusion}\label{sec:Conclusion}

We have argued both qualitatively and quantitatively that photorefractive
detection setups should generally see a fast signal at the second
harmonic (or more generally, at sum frequencies) of the ultrasound
waves in the tissue, which, like the slow (compared to the ultrasound frequency) envelope-related signal,
arises at the lowest order of perturbation theory and adds coherently
over all the collected speckles. Compared to the previously-discussed
slow signal, the fast signal is different and complementary. In terms
of spatial sensitivity, the fast signal measures high-spatial-frequency
Fourier components, and thus is blind to large structures but sensitive
to small features, fine texture, and sharp edges, while the slow signal is the opposite (see Fig.~\ref{fig:Pulses} for a comparison at 1MHz; the fast signal is sensitive to sub-millimeter structures while the slow signal is sensitive to several-millimeter structures and larger, along the direction of ultrasound propagation).
In terms of signal frequency, the fast signal requires a faster photodetector
and ADC (Nyquist rate of $4\times$ the highest ultrasound frequency), but should benefit from lower background, and more importantly
the same apparatus should be able to measure both the fast and slow
frequency bands simultaneously. In terms of contrast, the slow signal
effectively measures the sum $\left(M_{n}+\frac{1}{Z_{\text{US}}^{2}\omega_{\text{US}}^{2}}M_{d}\right)$
while the fast signal measures the difference $\left(M_{n}-\frac{1}{Z_{\text{US}}^{2}\omega_{\text{US}}^{2}}M_{d}\right)$.
These two will often be effectively the same (if $M_{n}\gg\frac{1}{Z_{\text{US}}^{2}\omega_{\text{US}}^{2}}M_{d}$,
as when $\lambda_{\text{US}}$ is less than twice the mean free path for photon scattering \cite{wang2001mechanisms,wang2001mechanisms2},
as is typical in tissues). But in certain cases, the difference term
offers intriguing sensing possibilities. For example, with low-frequency
ultrasound, it may be possible to arrange for $M_{n}\approx\frac{1}{Z_{\text{US}}^{2}\omega_{\text{US}}^{2}}M_{d}$
throughout the translucent tissue, but $M_{n}\gg\frac{1}{Z_{\text{US}}^{2}\omega_{\text{US}}^{2}}M_{d}$
in a particular area with unusually little light scattering, such
as a cyst full of relatively clear fluid. The high-frequency signal
component should then show a bright cyst on a dark background.

In future work, we plan to experimentally test and explore the predictions
herein.

{\bf Acknowledgement} I thank all those who generously offered feedback and criticism of this work, especially (but not exclusively) Joseph Hollmann, Charles DiMarzio, Krish Kotru, and Jeffrey Korn.

%
%

%

\bibliography{mybib}

 

\end{document}